\documentclass[epj]{svjour}
\usepackage{graphics}
\usepackage{amsmath}
\DeclareMathOperator\arctanh{arctanh}

\newcommand{\be}{\begin{equation}}
\newcommand{\ee}{\end{equation}}
\newcommand{\bea}{\begin{eqnarray}}
\newcommand{\eea}{\end{eqnarray}}

\begin{document}
\title{Emergence of multiverse in third quantized varying constants cosmologies}
%\subtitle{Do you have a subtitle?\\ If so, write it here}
\author{Adam Balcerzak\inst{1,2} \and Konrad Marosek\inst{3}% etc
% \thanks is optional - remove next line if not needed
%\thanks{\emph{Present address:} Insert the address here if needed}%
}                     % Do not remove
%
%\offprints{}          % Insert a name or remove this line
%
\institute{University of Szczecin,  Wielkopolska 15, 70-451 Szczecin,  Poland \and Copernicus Center for Interdisciplinary Studies, Szczepa\'nska 1/5, 31-011 Krak\'ow, Poland \and  Chair of Physics, Maritime University, Wa{\l }y Chrobrego 1-2, 70-500 Szczecin, Poland}
\date{Received: date / Revised version: date}
% The correct dates will be entered by Springer
%
\abstract{
Although the standard cosmological model explains most of the observed phenomena it still struggles with the problem of initial singularity. An interesting scenario in which the problem of the initial singularity is somehow circumvented was proposed in the context of string theory where the canonical quantisation procedure was additionally applied \cite{Birth}. A similar effect can be achieved in the context of the canonically quantized theory with varying speed of light and varying gravitational constant where both quantities are represented by non-minimally coupled scalar fields \cite{Balcerzak}. Such theory contains both the pre-big-bang contracting phase and the post-big-bang expanding phase and predicts non-vanishing probability of the transition from the former to the latter phase. In this paper we quantize such a theory once again by applying the third quantization scheme and show that the resulting theory contains scenario in which the whole multiverse is created from nothing. The generated family of the universes is described by the Bose-Einstein distribution.
 % end of PACS codes
} %end of abstract
\maketitle
\section{Introduction}
\label{intro}
The standard cosmological model based on  general theory of relativity and standard model of particle physics encounters a certain number of the fundamental problems. However, although it explains a vast majority of the observational data, it is  plagued with huge fine-tuning problems. Some of those can be resolved in the standard inflationary paradigm \cite{Linde}. Yet the classical theory of inflation is far from being complete since it does not provide initial conditions for the inflation to occur in a desired way. The incompleteness of this picture stretches to even more profound issues such as the occurrence of the initial singularity, the related questions about the beginning of time or the existence of epochs preceding big-bang. A consistent framework for investigating the above-mentioned questions is provided by string theory. The scenarios involving eras preceding big-bang singularity naturally arise in the cosmological models based on the string theory \cite{Gasperini}. The cosmological scenarios based on the tree level low energy effective action classically include both pre-big-bang and post-big-bang evolution phases separated by a singularity in which the curvature and the coupling reach infinite values. Employing the quantum cosmology Wheeler-DeWitt approach to the description of the near singularity regime gives rise to the scenario in which the universe passes form the pre-big-bang to the post-big-bang phase in a process that can be viewed as reflection (scattering) of the Wheeler-DeWitt wave function on  one dimensional exponential potential barrier in the minisuperspace \cite{Birth}. A similar scenario in which the universe scatters quantum mechanically over the singularity to finally enter the standard post-big-bang expansion era arises in the context of gravity theories with varying speed of light (VSL) and varying gravitational constant \cite{Balcerzak}. Many different VSL theories have been investigated in the literature so far, however, inventing a consistent picture in which the speed of light is allowed to vary encounters conceptual problems. One of these is the violation of the Lorentz invariance in case the speed of light is assumed to be dependent on the space-time coordinates \cite{Albrecht,Barrow1,Magueijo1,Clayton,Drummond,Clayton2}. Violation of the Lorentz invariance forces in turn an introduction of the preferred reference frame in which the particular mathematical structure of an investigated VSL theory is formulated. An interesting realisation of the mentioned above concepts is the VSL theory proposed in \cite{Albrecht,Barrow1}. The model additionally assumes that the degree of freedom representing speed of light is minimally coupled to the matter and the gravitational field in the preferred frame. Consequently, the equations of motion of such VSL gravity theory are identical to the standard Einstein equations with the speed of light being merely replaced by some time dependent functions. A different VSL gravity theory can be obtained by releasing the assumption of the minimal coupling \cite{Magueijo1}. The resulting set of equations of motion in such theory gains additional dynamical equation governing the evolution of the non-minimally coupled degree of freedom representing the speed of light. Another interesting example is VSL theory in which the speed of gravitons is assumed to be different from the speed of massless matter particles \cite{Clayton,Drummond,Clayton2}. This is achieved by introducing two different metrics. The first one describes the space-time geometry, the second one couples to the matter. The separate group of theories comprises VSL models which include corrections to the dispersion relation \cite{Rainbow}. Such corrections become relevant for the energy scales comparable with the Planck scale. The resulting group velocity of light depends on the energy scale.\\
\indent
In paper \cite{Balcerzak} it was shown that the near curvature singularity regime can be tackled within the framework of the theory which assumes that both the speed of light and the gravitational constant can vary. Both quantities are represented by scalar fields non-minimally coupled to the gravitational field in the preferred frame defined by the FLRW metric. This is different form the approach presented in \cite{Leszczynska} where the similar regime is investigated in the context of the model in which the degrees of freedom representing varying speed of light and gravitational constant minimally couple to gravity and matter.\\
\indent
The so-called third quantization is based on the formal similarity between the Wheeler-DeWitt and the Klein-Gordon equations \cite{Strominger,Robles}. The role of the Klein-Gordon field is played by the wave function in the Wheeler-DeWitt equation, which as a result of the third quantization becomes an operator acting on the Hilbert space. The third quantization itself is completely analogous to the quantization of the Klein-Gordon field. The resulting Hilbert space is spanned by an orthonormal basis which elements represent  occupation with universes which properties are determined by appropriate quantum numbers (these for example can be the momenta in the minisuperspace).\\
\indent
Our paper is organised as follows. In Sec. \ref{sec:1}  we quote the main results of the paper \cite{Balcerzak} which will be a starting point for the realization of the central task of our work consisting in elaboration of the scenario in which the whole multiverse emerges out of vacuum (``out of nothing''). In Sec. \ref{sec:2} we fulfill our assumed task by applying the third quantization procedure to the canonically quantized non-minimally coupled varying constants model and  by showing that the resulting theory includes a scenario in which the family of universes described by Bose-Einstein distribution is created out of nothing.

\section{Non-minimally coupled  varying $c$ and $G$ theories}
\label{sec:1}
\begin{figure}
\begin{center}
\resizebox{0.4\textwidth}{!}{\includegraphics{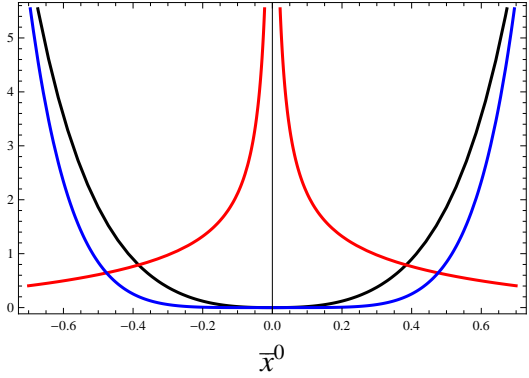}}
\caption{\label{acg} The scale factor $a$ (black), the speed of light $c$ (red) and the gravitational constant $G$ (blue) before ($\bar{x}^0<0$) and after ($\bar{x}^0>0$) the big-bang (curvature singularity).}
\end{center}
\end{figure}

The classical action defining the considered varying $c$ and $G$ model can be obtained by replacing each constant in the ordinary Einstein-Hilbert action with a certain function of a dynamical scalar degree of freedom. In order to keep  $c$ and $G$ positive during the cosmological evolution we link them to the newly introduced scalar degree of freedom  $\phi(x^\mu)$ and $\psi(x^\mu)$  by the following exponentials  $c^3=e^{\phi}$ and $G=e^\psi$. The resulting action reads \cite{Balcerzak}:
\begin{equation}
\label{action}
S=\int \sqrt{-g}  \left(\frac{e^{\phi}}{e^{\psi}}\right) \left[R+\Lambda + \omega (\partial_\mu \phi \partial^\mu \phi + \partial_\mu \psi \partial^\mu \psi)\right] d^4x,
\end{equation}
where $\omega$ is the parameter of the model. By application of the field redefinitions of the form:
\begin{equation}
\label{fred}
\phi = \frac{\beta}{\sqrt{2\omega}}+\frac{1}{2} \ln \delta,
\qquad
\psi = \frac{\beta}{\sqrt{2\omega}}-\frac{1}{2} \ln \delta,
\end{equation}
the action (\ref{action}) can be transformed into the form of Brans-Dicke action:
\bea
\label{actionBD}
S=\int \sqrt{-g}\left[ \delta (R+\Lambda) +\frac{\omega}{2}\frac{\partial_\mu \delta \partial^\mu \delta}{\delta} + \delta \partial_\mu \beta \partial^\mu \beta\right]d^4x.
\eea
The assumed dependence of $c$ on  space-time coordinates breaks the invariance of the action (\ref{action}) or (\ref{actionBD}) under general transformations of coordinates. This means that we have to specify the reference frame in which the action  given by (\ref{action}) or (\ref{actionBD}) is assumed to describe our varying $c$ and $G$ model. A natural step is to identify the above-mentioned  preferred reference frame with the cosmological frame defined by the flat FLRW metric:
\begin{equation}
\label{metric}
ds^2=-N^2 (dx^0)^2+a^2\left(dr^2+r^2 d\Omega^2\right),
\end{equation}
where $N$ is the lapse function and $a$ is the scale factor both depending on the parameter $x^0$ . Inserting the metric (\ref{metric}) into (\ref{actionBD}) gives the form of the action of our model in the cosmological frame:
\begin{eqnarray}
\label{action_sym} \nonumber
S &=& \frac{3 V_0}{8 \pi} \int dx^0 \left(-\frac{a^2}{N} a' \delta' - \frac{\delta}{N} a a'^2   + \Lambda \delta a^3 N  \right. \\
&-& \left.\frac{\omega}{2} \frac{a^3}{N} \frac{\delta'^2}{\delta}-\frac{a^3}{N}\delta \beta'^2 \right),
\end{eqnarray}
where $()'\equiv \frac{\partial}{\partial x^0}$.
A similar approach was introduced in papers \cite{Albrecht,Barrow1} where the VSL action and the corresponding variational principle was formulated in the cosmological frame. The solution to the model given by action (\ref{action_sym}) expressed in the gauge determined by $N=a^3\delta$ is:
\bea
\label{rozwio1}
a&=& \frac{1}{D^2 {(e^{ F x^0})}^2 \sinh ^ M |\sqrt{(A^2-9)\Lambda }x^0| },\\
\label{rozwio2}
\delta &=& \frac{D^6 {(e^{ F x^0})}^6}{\sinh ^ W |\sqrt{(A^2-9)\Lambda }x^0|},
\eea
where   $A=\frac{1}{\sqrt{1-2\omega}}$, $M=\frac{3-A^2}{9-A^2}$, $W=\frac{2A^2}{9-A^2}$ and $D$ and $F$ are some integration constants and the variable $x^0$
is the following function of the rescaled proper time $\bar{x}^0$ defined with its differential $d\bar{x}^0\equiv c(\bar{x}^0) d\tau$  with $\tau$ being here the proper time encountered by the comoving observer (for the detailed derivation of the solution given by (\ref{rozwio1}), (\ref{rozwio2}) and (\ref{conect}) see \cite{Balcerzak}):
\begin{equation}
\label{conect}
\begin{split}
x^0 &= \frac{2}{\sqrt{(A^2-9)\Lambda}}  \arctanh \left(e^{\sqrt{(A^2-9)\Lambda}\bar{x}^0}\right)\,,
\hspace{0.2cm}
\text{for $\bar{x}^0<0$}\,,
\\
x^0 &= \frac{2}{\sqrt{(A^2-9)\Lambda}}  \arctanh \left(e^{- \sqrt{(A^2-9)\Lambda}\bar{x}^0}\right)\,,
\hspace{0.2cm}
\text{for $\bar{x}^0>0$}\,,
\end{split}
\end{equation}
where as in  \cite{Balcerzak} we restrict our considerations to the models with  $A^2>9$. Fig. \ref{acg} depicts  qualitative behaviour of the scale factor $a$, the speed of light $c$ and the gravitational constant $G$ in the near high-curvature regime for the pre-big-bang  $\bar{x}^0<0$ and post-big-bang $\bar{x}^0>0$ phase. We see that as the universe approaches the curvature singularity ($a\rightarrow0$) the speed of light $c$ goes to infinity while the gravitational constant $G$ tends to zero. It means that the transition from pre-big-bang to post-big-bang phase occurs in the Newtonian limit. A similar behaviour of the scale factor occurs in the ekpyrotic \cite{Khoury,Khoury2}  and cyclic scenarios \cite{Steinhardt,Steinhardt2} where both pre-big-bang and post-big-bang eras are separated by the curvature singularity induced by vanishing scale factor. A qualitatively different evolution is encountered in the case of pre-big-bang scenarios based on the low-energy effective action of the string theory where curvature singularity is reached during the phase of accelerated expansion  \cite{Birth,Gasperini}.

Studying the quantum mechanical nature of the considered model requires analysis of the solutions of the  Wheeler-DeWitt equation. In order to obtain the Wheeler-DeWitt equation corresponding to our model first we have to  find the form of the  hamiltonian. By application of the following field transformation $X = \ln(a \sqrt{\delta})$, $Y = \frac{1}{2A} \ln \delta$, $I=AY-3X$, $J=3Y-AX$ and $B=\sqrt{\tilde{V}_0}\beta$ the action (\ref{action_sym}) can be reduced to the following form:
\begin{eqnarray}
\label{action final}
S&=& \int dx^0 \left[m (J'^2-I'^2) + \bar{\Lambda} e^{-2I}-B'^2\right].
\end{eqnarray}
The hamiltonian corresponding to our model is given by:
\begin{eqnarray}
\label{ham}
H&=&\frac{1}{4}\left[\frac{1}{m}\left(\pi_J^2- \pi_I^2\right)- \pi_B^2\right] -\bar{\Lambda}e^{-2I}.
\end{eqnarray}
Here $\bar{\Lambda} = \tilde{V}_0\Lambda$ with $\tilde{V}_0 = \frac{3 V_0}{8\pi}$ while $\pi_I=-2mI'$, and $\pi_B=-2B'$  are  the canonically conjugated momenta with $m = \frac{\tilde{V}_0}{9-A^2}$. The form of the hamiltonian proves that $\pi_J$ and $\pi_B$ are constant during the evolution. Therefore the classical evolution of the universe in the near curvature singularity regime can be reduced to the process of scattering of a particle on the exponential potential barrier. The solutions (\ref{rozwio1}) and (\ref{rozwio2}) expressed in term of the new variables $B$, $I$ and $J$ are given by:
\bea
\label{ham sol1}
B&=&-\frac{\pi_B}{2} x^0+ P, \\
\label{ham sol2}
I&=& \ln \sinh|\sqrt{(A^2-9)\Lambda }x^0|, \\
\label{ham sol3}
J&=&\frac{\pi_J}{2m} x^0+ E~,
\eea
where $E$ and $P$ are some constants.
\begin{figure}
\begin{center}
\resizebox{0.4\textwidth}{!}{%
  \includegraphics{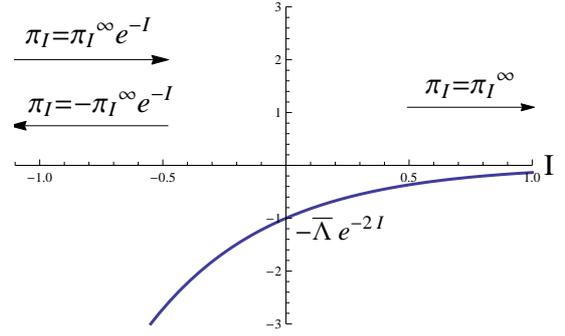}
}
\caption{ \label{bariera} Scattering on the exponential potential barrier. The curvature singularity occurs for $I\rightarrow \infty$. The pre-big-bang and post-big-bang branches in the low curvature limit are represented by the plane waves moving in the region located at $I\rightarrow -\infty$ characterized by momenta $\pi_I = \pi^\infty_I e^{-I}$ and $\pi_I = -\pi^\infty_I e^{-I}$ respectively.}
\end{center}
\end{figure}
By analysing the classical solutions (\ref{ham sol1})-(\ref{ham sol3}) we find that the near curvature singularity regime corresponds to the region located at $I\rightarrow \infty$. On the other hand the low-curvature regime for both the pre-big-bang and the post-big-bang branches is associated with the region located at  $I\rightarrow -\infty$.
%Equivalently the the near curvature singularity regime occurs for $\bar{x}^0\rightarrow0$, while the low-curvature regime is given by %$\bar{x}^0\rightarrow-\infty$ and $\bar{x}^0\rightarrow\infty$ for the pre-big-bang and the post-big-bang branch respectively.
Both regimes correspond to the particular asymptotic values of the momentum $\pi_I$. In the high-curvature (near big-bang) limit $I\rightarrow \infty$ we have
  \[ \pi_I = \left\{
  \begin{array}{l l}
    \pi^\infty_I & \quad \text{collapsing pre-big-bang }\\
    -\pi^\infty_I & \quad \text{expanding post-big-bang}
  \end{array} \right.\]
while in the low-curvature (far away from big-bang) limit $I\rightarrow -\infty$  the momentum $\pi_I$ is given by:
   \[ \pi_I = \left\{
  \begin{array}{l l}
    \pi^\infty_I e^{-I} & \quad \text{collapsing pre-big-bang}\\
    -\pi^\infty_I e^{-I} & \quad \text{expanding post-big-bang}
  \end{array} \right.\]
where $\pi^\infty_I\equiv 2 \tilde{V}_0\sqrt{\frac{\Lambda}{A^2-9}}$ (again for the detailed derivation see \cite{Balcerzak}). In order to obtain the quantum theory we apply the Jordan quantization rules and replace the canonical momenta with the following operators: $\pi_J\rightarrow \hat{\pi}_J=-i \frac{\partial}{\partial J}$, $\pi_I\rightarrow \hat{\pi}_I=-i \frac{\partial}{\partial I}$ and $\pi_B\rightarrow \hat{\pi}_B= -i \frac{\partial}{\partial B}$. The corresponding Wheeler-DeWitt equation is:
\begin{equation}
\label{WDW}
\left\{\frac{1}{4}\left[\frac{1}{m}\left(\frac{\partial^2}{\partial I^2}-\frac{\partial^2}{\partial J^2}\right)+ \frac{\partial^2}{\partial B^2}\right] -\bar{\Lambda}e^{-2I}\right\}\Phi=0 \,
\end{equation}
and its separable solution is given by \cite{DabQ}:
\begin{equation}
\label{psi}
\Phi=\alpha(J)\gamma(B)\beta(I) \,,
\end{equation}
where
\begin{equation}
\alpha(J) = e^{i  k_1 r J}, \hspace{0.1cm} \gamma(B) = e^{i2 k_2 B}, \hspace{0.1 cm}
\label{bessel}
\beta(I) = J_{-i\pi^\infty_I}(\pi^\infty_I e^{-I}),
\end{equation}
where $r=2\sqrt{\frac{\tilde{V}_0}{A^2-9}}$ and
\be \label{hamcon}
\tilde{V}_0 \Lambda =  k_1^2+k_2^2.
\ee
Asymptotically for $I\rightarrow\infty$ the wave function $\Phi$ is an eigenfunction of the momentum operator $\hat{\pi}_I$ since $\hat{\pi}_I \beta(I)=\pi^\infty_I\beta(I).$ Thus, in the near curvature singularity regime  $\Phi$ represents only those modes which are associated with the collapsing classical solution \cite{Birth,DabKief}. On the other hand asymptotically in the low curvature regime  ($I \rightarrow -\infty$) the Bessel function $J_{in}$ can be written as $J_{in}(z)=\Psi_1+\Psi_2$ with $\Psi_1=\sqrt{\frac{1}{2 \pi z}} e^{i\left(z-\frac{\pi}{2}in - \frac{\pi}{4}\right)}$ and $\Psi_2=\sqrt{\frac{1}{2 \pi z}} e^{-i\left(z-\frac{\pi}{2}in - \frac{\pi}{4}\right)}$ where  $n\equiv-\pi^\infty_I$. Since both $\Psi_1$ and $\Psi_2$ are eigenfunctions of the momentum operator $\hat{\pi}_I$ to the eigenvalues $-\pi^\infty_I e^{-I}$ and $\pi^\infty_I e^{-I}$ respectively we find that asymptotically for $I \rightarrow -\infty$ the function $\Psi_2$ represents the pre-big-bang collapsing branch while the function $\Psi_1$ represents the post-big-bang expanding branch. Thus the transition form the pre-big-bang low curvature initial state to the post-big-bang low curvature final state corresponds to the stationary quantum mechanical scattering of the plane wave (representing the particle with the definite value of the momentum)  on the exponential potential barrier in the minisuperspace (see Fig. \ref{bariera}).

\section{Emergence of multiverse in the third quantized varying $c$ and $G$ model}
\label{sec:2}
In this section we will argue that the third quantized non-minimally coupled varying constants model includes a scenario in which the whole bunch of universes emerges out of the initial vacuum (containing no universes) in the process analogous to the particle creation in an evolving background. The idea of third quantization assumes that the multiverse can be treated as a many-particle system (with non conserved number of particles) where the individual particles with their trajectories in spacetime are replaced by the universes making up the multiverse existing in the minisuperspace \cite{Strominger,Robles}.  The Wheeler-DeWitt \eqref{WDW} can be rewritten in the form of Klein-Gordon equation:
\begin{equation}
\label{KG}
\Phi'' - \Delta \Phi + m_{eff}^2 \Phi=0,
\end{equation}
where $( )'=\frac{\partial}{\partial \eta}$ and $\Delta = \frac{\partial^2}{\partial x_1^2}+\frac{\partial^2}{\partial x_2^2}$, $m_{eff}^2= \overline{\Lambda}e^{-\frac{2}{r}\eta}$. The relation between the new variables $\eta$, $x_1$ and $x_2$ and the old variables $I$, $J$ and $B$ are:
\bea
\eta&\equiv& rI,\\
x_1&\equiv& rJ, \\
x_2&\equiv& 2B.
\eea
It is useful to define the following vectors: $\vec{x}\equiv(x_1,x_2)$, $\vec{k}\equiv(k_1,k_2)$. The formal analogy between Wheeler-DeWitt  and the Klein-Gordon equations allows us to \emph{third} quantize the wave function $\Phi$ by \emph{formally} applying the standard quantization procedure of the quantum field theory. Following this procedure we promote the wave function $\Phi$ to be an operator $\hat{\Phi}$ acting on an associated quantum space of state \cite{mukhanov}. The expansion of the field operator $\hat{\Phi}$  in terms of mode functions $v_k(\eta)$ is:
\begin{equation}
\label{exp1}
\hat{\Phi}(\vec{x},\eta)=\frac{1}{\sqrt{2}}\int \frac{d^2k}{2\pi}[e^{i\vec{k}\cdot\vec{x}}v_k^*(\eta)\hat{a}^-_{\vec{k}}+e^{-i\vec{k}\cdot\vec{x}}v_k(\eta)\hat{a}^+_{\vec{k}}],
\end{equation}
where $k\equiv|\vec{k}|$. The mode functions $v_k(\eta)$ fulfil the mode equation
\be \label{modeeq}
v_k(\eta)''+\omega_k(\eta)^2 v_k(\eta)=0,
\ee
where $\omega_k(\eta)=\sqrt{k^2+m_{eff}^2(\eta)}$, and the normalisation condition
\be
W(v_k(\eta),v^*_k(\eta))=2i,
\ee
where $W$ is a Wronskian. We impose the standard commutation rules on the creation an annihilation operators $\hat{a}^-_{\vec{k}}$ and $\hat{a}^+_{\vec{k}}$:
\begin{eqnarray}
\label{commut}
[\hat{a}^-_{\vec{k}},\hat{a}^+_{\vec{k}'}]&=&\delta(\vec{k}-\vec{k}'), \\
\lbrack \hat{a}^-_{\vec{k}},\hat{a}^-_{\vec{k}'}\rbrack &=& 0,\\
\lbrack \hat{a}^+_{\vec{k}},\hat{a}^+_{\vec{k}'}\rbrack &=& 0.
\end{eqnarray}
Instead of $v_k(\eta)$ any linear combinations of the form
\be
u_k(\eta)=\alpha_k v_k(\eta)+\beta_k v^*_k(\eta),
\ee
with the coefficients $\alpha_k$ and $\beta_k$ fulfilling the normalisation condition $|\alpha_k|^2-|\beta_k|^2=1$ could be used as the mode functions since any such linear combination is also a solution of the mode equation (\ref{modeeq}). The field operator $\hat{\Phi}$ expressed in terms of the mode  functions $u_k(\eta)$ is given by:
\begin{equation}
\label{exp2}
\hat{\Phi}(\vec{x},\eta)=\frac{1}{\sqrt{2}}\int \frac{d^2k}{2\pi}[e^{i\vec{k}\cdot\vec{x}}u_k^*(\eta)\hat{b}^-_{\vec{k}}+e^{-i\vec{k}\cdot\vec{x}}u_k(\eta)\hat{b}^+_{\vec{k}}]~,
\end{equation}
where $\hat{b}^-_{\vec{k}}$ and $\hat{b}^+_{\vec{k}}$ is a different set of the creation and annihilation operators satisfying the standard commutations relations given by:
\begin{eqnarray}
\label{commut2}
[\hat{b}^-_{\vec{k}},\hat{b}^+_{\vec{k}'}]&=&\delta(\vec{k}-\vec{k}'), \\
\lbrack \hat{b}^-_{\vec{k}},\hat{b}^-_{\vec{k}'}\rbrack &=& 0,\\
\lbrack \hat{b}^+_{\vec{k}},\hat{b}^+_{\vec{k}'}\rbrack &=& 0.
\end{eqnarray}
By comparing the two expansions (\ref{exp1}) and (\ref{exp2})  we can obtain the Bogolyubov transformations in the form:
\begin{eqnarray}
\hat{a}^-_{\vec{k}}&=&\alpha_k^*\hat{b}^-_{\vec{k}}+\beta_k\hat{b}^+_{-\vec{k}}, \\
\hat{a}^+_{\vec{k}}&=&\alpha_k \hat{b}^+_{\vec{k}}+ \beta^*_k  \hat{b}^-_{-\vec{k}}.
\end{eqnarray}
The explicit values of the Bogolyubov coefficients $\alpha_k$ and $\beta_k$ are:
\begin{eqnarray}
\alpha_k&=&\frac{W(u_k,v^*_k)}{2i}, \\
\beta_k&=&\frac{W(v_k,u_k)}{2i}.
\end{eqnarray}
We define the two different vacuum states in a standard way:
\begin{eqnarray}
\hat{a}^-_{\vec{k}}|_{(a)}0\rangle&=&0, \\
\hat{b}^-_{\vec{k}} |_{(b)}0\rangle&=&0.
\end{eqnarray}
Following the paper \cite{Pimentel} we interpret both vacuum states $|_{(a)}0\rangle$ and $|_{(b)}0\rangle$ as  state vectors which represent the ``ground states'' of the multiverse, where the notion  ``ground state'' refers to the ``empty'' multiverse (a multiverse containing no ``a-universes'' for $|_{(a)}0\rangle$ vector and a multiverse containing no ``b-universes'' for $|_{(b)}0\rangle$  vector). The creation and annihilation operators can be used to build the two sets of the excited state of the multiverse. Thus the quantum state of the multiverse containing ``a-universes'' with $m$ universes in  mode $\vec{k}_1$,  $n$ universes in  in mode $\vec{k}_2$, etc. is represented by
\be
|_{(a)}m_{\vec{k}_1},n_{\vec{k}_2},...\rangle \equiv \frac{1}{\sqrt{m!n!...}}\lbrack (\hat{a}^+_{\vec{k}_1})^m (\hat{a}^+_{\vec{k}_2})^n ...\rbrack |_{(a)}0\rangle,
\ee
while the quantum state of the multiverse containing ``b-universes'' with $m$ universes in  mode $\vec{k}_1$,  $n$ universes in  in mode $\vec{k}_2$, etc. is represented by
\be
|_{(b)}m_{\vec{k}_1},n_{\vec{k}_2},...\rangle \equiv \frac{1}{\sqrt{m!n!...}}\lbrack (\hat{b}^+_{\vec{k}_1})^m (\hat{b}^+_{\vec{k}_2})^n ...\rbrack |_{(b)}0\rangle.
\ee
Thus we have built an orthonormal basis on the Hilbert space of the multiverse. This means that any arbitrary vector representing  state of the multiverse can be written as a linear combination of the excited states:
\bea
|\psi\rangle &=&\sum_{m,n,...} C^{(a)}_{mn...}|_{(a)}m_{\vec{k}_1},n_{\vec{k}_2},...\rangle= \\ \nonumber
&=&\sum_{m,n,...} C^{(b)}_{mn...}|_{(b)}m_{\vec{k}_1},n_{\vec{k}_2},...\rangle.
\eea
The two vacuum states $|_{(a)}0\rangle$ and $|_{(b)}0\rangle$ are generally different vectors. This can be seen by calculating the expectation value of the ``a-universe'' number operator in the vacuum state $|_{(b)}0\rangle$:
\begin{equation}
\langle_{(b)}0|\hat{N}^{(a)}_{\vec{k}}|_{(b)}0\rangle=\langle_{(b)}0| \hat{a}^+_{\vec{k}} \hat{a}^-_{\vec{k}}|_{(b)}0\rangle=|\beta_k|^2\delta^{(3)}(0)~.
\end{equation}
The divergent factor $\delta^{(3)}(0)$ accounts for an infinite spatial volume and hence the mean density of ``a-universes'' in the mode $\vec{k}$ is:
\begin{equation}
n_{\vec{k}}=|\beta_k|^2~.
\end{equation}

Now, we are going to introduce a scenario in which the whole family of the universes is created from nothing. Let us first assume that initially the vector representing the quantum  state of the multiverse is described by the ``b-vacuum'' state  $|_{(b)}0\rangle$ which is completely specified by the set of mode functions
\be
u_k=A J_{-ikr}(x)
\ee
that solve the mode equation (\ref{modeeq}) in the high-curvature limit (which appear  for $\eta\rightarrow\infty$), where $A$ is the normalisation constant and $x\equiv\sqrt{\bar{\Lambda}} e^{-\eta/r}$. Since the structure of the vacuum state is controlled by the instantaneous value of the background curvature the initial vacuum state is identical with $|_{(b)}0\rangle$. Thus, initially the multiverse contains no universes at all. Due to stationarity of the scattering process the instantaneous quantum state of the multiverse does not evolve, so the final state of the multiverse (in the low-curvature limit that occurs for $\eta\rightarrow-\infty$) is still represented by $|_{(b)}0\rangle$. Due to the variation of the background curvature the notion of the vacuum is not invariant and the final vacuum state corresponding to the low-curvature limit (for $\eta\rightarrow-\infty$) is identical with $|_{(a)}0\rangle$ which, on the other hand, is completely specified by the set of mode functions
\be
v_k=B H^{(2)}_{-ikr}(x)
\ee
being the solution of the mode equation (\ref{modeeq}) in the low-curvature limit (for $\eta\rightarrow-\infty$),  where $B$ is the normalisation constant and $H^{(2)}_{-ikr}(x)$ is the Hankel function of the second kind. Therefore, the quantity:
\be \label{distri}
n_{\vec{k}}=|\beta_k|^2=\left|\frac{W(v_k,u_k)}{2i}\right|^2
\ee
can be interpreted  as the average number of the universes created from nothing described with the quantum number $\vec{k}$. By calculating the Wronskian in (\ref{distri}) we obtain that:
\begin{equation}
\label{avnumb}
n_{\vec{k}}=\frac{1}{e^{2 \pi k r}-1}~.
\end{equation}
The formula (\ref{avnumb}) describes the Bose-Einstein distribution for the temperature $T=\frac{1}{2 \pi k_B}$, where $k_B$ is the Boltzmann constant provided that the energy of the bosons is identified with the quantity $kr$. Taking into account the eq. (\ref{hamcon}) we obtain that $kr=\pi^{\infty}_I\equiv 2 \tilde{V}_0 \sqrt{\frac{\Lambda}{A^2-9}}$. Thus, the distribution (\ref{avnumb}) expressed in terms of the cosmological constant  $\Lambda$ reads:
\begin{equation}
\label{avnumb2}
\frac{\pi\tilde{V}_0}{e^{4 \pi \tilde{V}_0 \sqrt{\frac{\Lambda}{A^2-9}}}-1}~.
\end{equation}
The formula (\ref{avnumb2}) expresses  the average number  of the  universes created from nothing with the value of the cosmological constant $\Lambda$ in the interval $(\Lambda,\Lambda+d\Lambda)$. From (\ref{avnumb2}) we can see that the concentration of universes characterised by small value of the vacuum energy density is large. On the contrary, for larger values of the vacuum energy density the concentration of universes tends to zero.

\section{Discussion}
\label{sec:3}
The canonically quantized non-minimally coupled varying $c$ and $G$ theory includes a scenario in which the transition from the pre-big-bang contraction to the post-big-bang expansion occurs as a consequence of the scattering of the plane wave on the exponential potential barrier in the minisuperspace. This is similar to the scenarios included in the string cosmologies \cite{Birth}. The third quantization  (a procedure analogous to the quantization of the Klein-Gordon field) of such a theory leads to the scenario in which the whole bunch of universes is created out of vacuum. The third quantization scheme was already applied to discuss the transition from expanding to contracting cosmological phases (and vice-versa) also in \cite{Buonanno,Gasperini3}. The Hilbert space that emerges in the process of third quantization comprising all the states of the multiverse can be used as a base for introducing some standard notions of the ordinary quantum mechanics that exploit the linear structure of the space of quantum states. Particularly important here is the concept of quantum entanglement which relies on the notion of the tensorial product of the Hilbert spaces associated with quantum subsystems making up the whole physical setup. The scenarios in which the components of the multiverse experience their mutual presence via quantum entanglement was considered in \cite{Robles2,Robles3}. Concepts of that kind can constitute a basis for introducing models in which the quantum entanglement between different universes influences the cosmological observables making the idea of multiverse observationally testable \cite{Kraemer,Kraemer2}.


\begin{thebibliography}{}

\bibitem{Linde}
A. D. Linde,  Particle physics and inflationary cosmology. Harwood: New York, USA, 1990.

\bibitem{Birth}
M. Gasperini, G. Veneziano,  Birth of the Universe as quantum scattering in string cosmology. {\em Gen. Rel. Grav.} {\bf 1996}, {\em  28}, 1301–1307.

\bibitem{Gasperini}
M. Gasperini, G. Veneziano,  The pre-big bang scenario in string cosmology. {\em Phys. Rep.} {\bf  2003}, {\em  373}, 1–212.

\bibitem{Balcerzak}
A. Balcerzak,  Non-minimally coupled varying constants quantum cosmologies. {\em JCAP} {\bf  2015}, {\em  04}, 019.

\bibitem{Albrecht}
A. Albrecht, J. Magueijo,    Time varying speed of light as a solution to cosmological puzzles. {\em Phys. Rev. D} {\bf  1999}, {\em  59}, 043516.

\bibitem{Barrow1}
J. D. Barrow,  Cosmologies with varying light speed. {\em Phys. Rev. D} {\bf  1999}, {\em  59}, 043515.

\bibitem{Magueijo1}
J. Magueijo, Covariant and locally Lorentz-invariant varying speed of light theories. {\em Phys. Rev. D} {\bf  2000}, {\em  62}, 103521.

\bibitem{Clayton}
M. A. Clayton, J. W. Moffat, Dynamical mechanism for varying light velocity as a solution to cosmological problems. {\em Phys. Lett. B} {\bf  1999}, {\em  460}, 263.

\bibitem{Drummond}
I. T. Drummond, Bimetric gravity and ``dark matter''. {\em Phys. Rev. D} {\bf  2001}, {\em  63}, 043503.

\bibitem{Clayton2}
M. A. Clayton, J. W. Moffat, Scalar-Tensor Gravity Theory For Dynamical Light Velocity {\em Phys. Lett. B} {\bf  2000}, {\em  477}, 269.

\bibitem{Rainbow}
J. Magueijo, L. Smolin,   Gravity's Rainbow. {\em Class. Quant. Grav.} {\bf  2004}, {\em  21}, 1725.

\bibitem{Leszczynska}
K. Leszczy\'nska; M.P. D\c{a}browski, A. Balcerzak,  Varying constants quantum cosmology. {\em JCAP} {\bf  2015}, {\em  02}, 012.

\bibitem{Strominger}
A. Strominger,  Baby universes, in Quantum Cosmology and Baby Universes, Vol. 7, ed. by S. Coleman, J. B. Hartle, T. Piran and S. Weinberg, World Scientific, London (1990).

\bibitem{Robles}
S. Robles-P\'erez, P. F. Gonzalez-Diaz, Quantum state of the multiverse. {\em Phys. Rev. D} {\bf 2010}, {\em  81}, 083529.

\bibitem{Khoury}
J. Khoury, B. A. Ovrut, P. J. Steinhardt, N. Turok,  Ekpyrotic universe: Colliding branes and the origin of the hot big bang. {\em Phys. Rev. D} {\bf  2001}, {\em  64}, 123522.

\bibitem{Khoury2}
J. Khoury,  B. A. Ovrut, N. Seiberg, P. J. Steinhardt, N. Turok, From big crunch to big bang. {\em Phys. Rev. D} {\bf 2002}, {\em  65}, 086007.

\bibitem{Steinhardt}
P. J. Steinhardt, N. Turok,  Cosmic evolution in a cyclic universe. {\em Phys. Rev. D} {\bf 2002}, {\em  65}, 126003.

\bibitem{Steinhardt2}
P. J. Steinhardt, N. Turok,  A Cyclic Model of the Universe. {\em Science} {\bf 2002}, {\em  296}, 1436.

\bibitem{DabQ}
M.P. D\c{a}browski,  Quantum String Cosmologies. {\em Ann. Phys. (Leipzig)} {\bf 2001}, {\em  10}, 195.

\bibitem{DabKief}
M.P. D\c{a}browski, C. Kiefer,  Boundary Conditions in Quantum String Cosmology. {\em Phys. Lett. B} {\bf 1997}, {\em  397}, 185.



\bibitem{mukhanov}
V. Mukhanov, S. Winitzki, Introduction to quantum effects in gravity. Cambridge University Press: Cambridge, United Kingdom, 2007.

\bibitem{Pimentel}
L. O. Pimentel, C. Mora, Third quantization of Brans–Dicke cosmology. {\em Phys. Lett. A} {\bf 2001}, {\em  280}, 191.

\bibitem{Buonanno}
A. Buonanno, M. Gasperini, M. Maggiore and C. Ungarelli,
Expanding and contracting universes in third quantized string cosmology. {\em  Class. Quant. Grav.} {\bf 1997}, {\em 14}, L97.

\bibitem{Gasperini3}
M. Gasperini, Birth of the universe as antitunneling from the string perturbative vacuum. {\em Int. J. Mod. Phys. D} {\bf 2001}, {\em 10}, 15.


\bibitem{Robles2}
S. Robles-P\'erez, A. Alonso-Serrano, C. Bastos, O. Bertolami, Vacuum decay in an interacting multiverse.  {\em Phys. Lett. B} {\bf  2016}, {\em  759}, 328.

\bibitem{Robles3}
S. Robles-P\'erez, A. Balcerzak, M. P. D\c{a}browski, M. Kraemer, Interuniversal entanglement in a cyclic multiverse,  {\em Phys. Rev. D} {\bf 2017}, {\em  95} 083505.

\bibitem{Kraemer}
J. Morais, M. Bouhmadi-Lopez, M. Kraemer, S. Robles-P\'erez, Pre-inflation from the multiverse: can it solve the quadrupole problem in the cosmic microwave background?, {\em  Eur. Phys. J. C} {\bf 2018}, {\em  78}  240.

\bibitem{Kraemer2}
M. Bouhmadi-Lopez, M. Kraemer, J. Morais, S. Robles-P\'erez, The interacting multiverse and its effect on the cosmic microwave background,  {\em JCAP} {\bf 2019}, {\em  2019}, 057.

\end{thebibliography}
\end{document}